\documentclass[journal]{IEEEtran}
\usepackage{mathpazo}
\usepackage{times}

\usepackage[utf8]{inputenc} 
\usepackage[T1]{fontenc}
\usepackage{url}
\usepackage{ifthen}
\usepackage{cite}
\usepackage{graphicx}
\usepackage{amssymb}
\usepackage{xfrac}
\usepackage[cmex10]{amsmath} 
\usepackage{color}
\usepackage[ruled,vlined,linesnumbered]{algorithm2e}

\usepackage{nicefrac}

\usepackage{caption}
\newcommand{\code}{\ttfamily\bfseries}

\newcommand{\dfn}{\bfseries\itshape}

\renewcommand{\ge}{\geqslant}
\renewcommand{\geq}{\geqslant}

\newcommand{\be}[1]{\begin{equation}\label{#1}}
\newcommand{\ee}{\end{equation}} 
\newcommand{\eq}[1]{(\ref{#1})}

\newcommand{\Cref}[1]{Co\-ro\-lla\-ry\,\ref{#1}}
\newcommand{\Fref}[1]{Fi\-g\-u\-re\,\ref{#1}}
\newcommand{\Sref}[1]{Sec\-t\-ion\,\ref{#1}}

\newcommand{\cA}{{\cal A}}

\newcommand{\cF}{{\cal F}}

\DeclareMathAlphabet{\mathbfsl}{OT1}{ppl}{b}{it} 

\newcommand{\bT}{\mathbfsl{T}}

\newcommand{\ccc}{\mathbfsl{c}} 
\newcommand{\ddd}{\mathbfsl{d}}

\newcommand{\uuu}{\mathbfsl{u}} 
\newcommand{\vvv}{\mathbfsl{v}}
 
\newcommand{\xxx}{\mathbfsl{x}}
\newcommand{\yyy}{\mathbfsl{y}}

\newcommand{\Arikan}{Ar{\i}\-kan}

\newcommand{\shalf}{\mbox{\large{$\nicefrac{1}{2}$}}}

\DeclareMathOperator{\wt}{wt}
\renewcommand{\bT}{{\bf T}}

\begin{document}
\title{$\,$\\[-1.62ex] \textbf{List Decoding of Ar{\i}kan's PAC Codes\\[0.27ex]}}
\author{%
                  \textbf{Hanwen Yao,}
\IEEEauthorblockN{\textbf{Arman Fazeli,} 
        and       \textbf{Alexander Vardy}}\\[0.27ex]
\IEEEauthorblockA{University of California San Diego, La Jolla, CA 92093, USA\\}
{\code\{hwyao,avardy,afazelic\}@ucsd.edu\vspace*{-2.70ex}}%
\thanks{
To be presented in part
at the IEEE International Symp\-osium on Informat\-ion Theory in 
June 2020 (submitted for review January\,15, 2020).}
}

\maketitle

\begin{abstract}
Polar coding gives rise to the first explicit family~of codes that
provably achieve capacity with efficient encoding and decoding
for a wide range of channels. However, its performance at short 
block lengths under standard successive cancellation~de\-coding
is far from optimal.
A well-known way to improve~the~per-formance of polar codes at 
short block lengths is CRC~precod\-ing followed by successive-cancellation
list decoding. This approach, along with various refinements thereof,
has largely remained the state of the art in polar coding since it was 
introduced in 2011.\linebreak
Last year, Ar{\i}kan presented a new polar coding scheme, 
which~he called \emph{polarization-adjusted convolutional} (PAC)~\emph{codes}.
Such PAC codes provide another dramatic improvement in performance
as compared to CRC-aided list decoding. These codes~are~based~primarily
upon the following main ideas: replacing CRC~pre\-coding with 
{convolutional precoding} (under appropriate rate profiling) and 
replacing list decoding by \emph{sequential decoding}.
Ar{\i}kan's~simulation results show that PAC codes, resulting from 
the combina\-tion of these ideas, are quite close to finite-length 
lower bounds\linebreak
on the performance of \emph{any} code under ML decoding.\hspace*{12ex}

One of our main goals in this paper is to answer the following question:
is sequential decoding essential for the superior perfor\-mance of PAC codes? 
We show that similar performance can be achieved using list decoding 
when the list size $L$ is moderately large (say, $L \ge 128$).
List decoding has distinct advantages~over sequential decoding in certain
scenarios such as low-SNR regimes\linebreak
or situations where the \emph{worst-case}
complexity/latency is~the~prim\-ary constraint.
Another objective is to provide some insights~into the remarkable performance
of PAC codes. 
We first observe that both sequential decoding
and list decoding of PAC codes closely match ML decoding thereof.
We then estimate the number~of~low\linebreak
weight codewords in PAC codes, and use these estimates 
to appro- ximate the union bound on their performance under ML decod-ing.
These results indicate 
that PAC codes are superior to 
polar\linebreak codes and Reed-Muller codes, and suggest that the goal
of rate-profiling may be to optimize the weight distribution 
at low weights.~~\vspace{0.18ex}
\end{abstract}


\vspace{0.90ex}
\section{Introduction}
\label{sec:intro}

\noindent 
Polar coding, pioneered by Ar\i kan~\cite{Arikan09}, 
gives rise to the~first~explicit family of codes that provably achieve
capacity for~a~wide range of channels with efficient encoding and decoding.
However, it is well known that at short block lengths 
the~perform\-ance of polar codes
is far from optimal. 

\begin{figure}[t!]
\centering
$\,$\\[2.70ex]
\includegraphics[trim=60 220 60 220, width=3.15in]%
{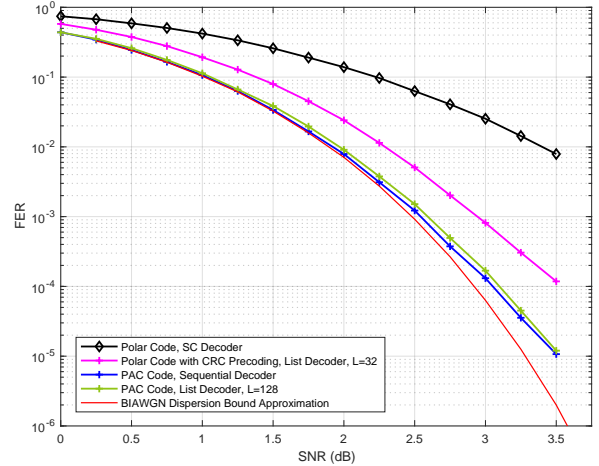}
\\[1.80ex]
\caption{Performance of PAC codes versus polar codes}
  \label{fig:performance}
\end{figure}

For example, the performance of a polar code of length~$128$ and rate $\shalf$
on the binary-input AWGN channel under~standard successive cancellation (SC) 
decoding is shown 
in Figure\,\ref{fig:performance}.
Figure\,\ref{fig:performance} largely reproduces the simulation~results
presented by \Arikan{} in \cite{Arikan19}. Codes 
of length~$128$ and rate~$\shalf$~serve 
as the running example throughout
\Arikan's recent paper~\cite{Arikan19}, and 
we will adopt this strategy herein. 
We make no attempt~to optimize these codes; rather, our goal is
to follow \Arikan~\cite{Arikan19} as closely as possible.
Also shown in Fig\-ure\,\ref{fig:performance} 
is the BIAWGN dispersion bound approximation 
for~such codes.
This 
can be thought of as an estimate of the perform\-ance 
of random codes under ML decoding (see \cite{PPV10}). 
Clearly, at length~$128$, there is a tremendous gap between 
polar codes under SC decoding and the best achievable performance.

As shown in~\cite{TV15} and other papers, the 
reasons for this~gap are two-fold: 
the polar code itself is weak at such short~lengths
and SC decoding is weak in comparison with ML decoding.\linebreak 
%
A~well-known 
way to address both problems
is CRC~precod\-ing followed by successive-cancellation
list (SCL) decoding.
Following~\cite{Arikan19},
the performance of CRC-aided polar codes (with $8$-bit CRC)~of
rate~$\shalf$ under SCL decoding with list-size $32$~is 
also shown in \Fref{fig:performance}.
%
%
This approach, along with various refinements~the\-reof
\cite{LST12,MT14,Niu12}, 
has largely remained the state of the art in polar coding since it was first
introduced in~\cite{TV15}.

In his Shannon Lecture at the ISIT in 2019, 
Erdal Ar{\i}kan~presented a significant breakthrough in polar
coding, which~boosts the performance of polar codes at short 
lengths. 
Specifically, Ar{\i}kan~\cite{Arikan19} proposed
a new polar coding scheme, 
which he~calls 
\emph{polarization-adjusted convolutional (PAC)}~\emph{codes}.
Remarkably, under sequential decoding, 
the performance of PAC codes is very close 
to the BIAWGN dispersion bound approx\-imation.
The performance of PAC codes of length $128$ and rate $\shalf$
is also shown (in blue and green) 
in Figure\,\ref{fig:performance}.~

\vspace{1.44ex}
\subsection{Brief Overview of PAC Codes}

\begin{figure}[t!] 
  \centering
\includegraphics[width=3.1in]{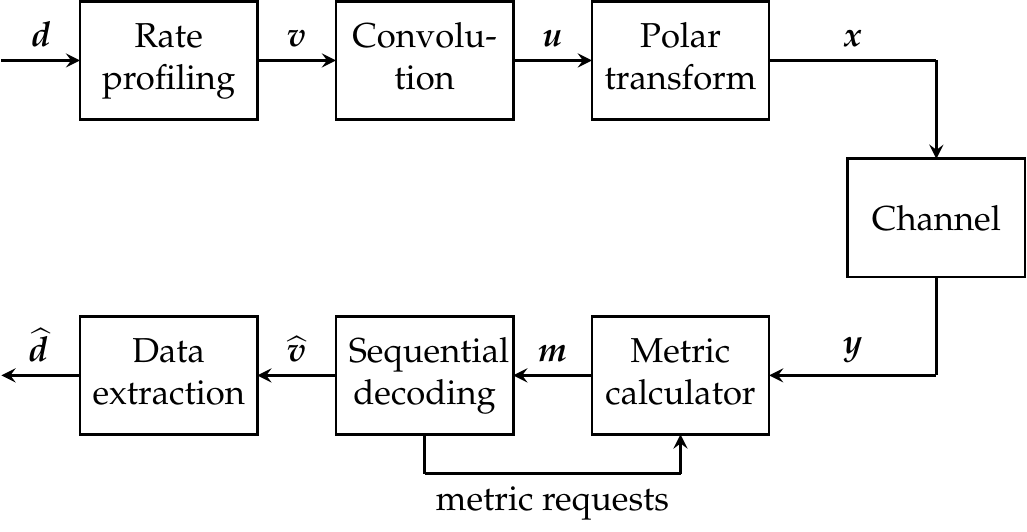}
\vspace{0.36ex}
\caption{PAC coding scheme, reproduced from~\cite{Arikan19}}
  \label{fig:PAC_scheme}
\vspace*{-0.90ex}
\end{figure}

\noindent
\Arikan's PAC codes~\cite{Arikan19} are based primarily
upon the following two innovations: replacing CRC~pre\-coding with 
\emph{convolutional precoding} (under appropriate rate-profiling, 
to be~dis\-cussed la\-ter) \pagebreak
and replacing list decoding by \emph{sequential decod\-ing}.
The encoding and decoding of PAC codes are shown schematically in 
Figure\,\ref{fig:PAC_scheme}, which is reproduced from~\cite{Arikan19}.

Referring to Figure\,\ref{fig:PAC_scheme}, let's consider
an $(n,k)$ PAC~code. On the encoding side, Ar{\i}kan uses 
a rate-$1$ convolutional precoder concatenated with a standard polar encoder.
Only $k$ out of the $n$ bits of the input $\vvv$ 
to the convolutional precoder carry the information (or data) vector $\ddd$.
The remaining $n-k$ bits of $\vvv$ are set to $0$. 
Just like for conventional polar codes, the overall 
performance 
crucially depends upon \emph{which positions} in $\vvv$ 
carry~information and which are frozen to $0$. This choice~of~frozen
positions in $\vvv$, \Arikan{} has termed \emph{\dfn rate-profiling}.
Unlike
conventional polar codes, the optimal rate-profiling choice is \emph{not} 
known. In fact, it is not even clear what optimization~criterion 
should govern this choice, although we hope to shed some light\linebreak 
on this in \Sref{sec:analysis}.

\looseness=-1
The main operation on the decoder side is sequential decoding.
Specifically, Ar{\i}kan employs 
Fano decoding of the~convolutional code to estimate its input $\vvv$. 
The path metrics used by this sequential decoder are obtained via
repeated calls to the successive-cancellation decoder for the underlying
polar code.

\vspace{1.08ex}
\subsection{Our Contributions}

\noindent
\looseness=-1
One of our main goals in this paper is to answer the following question:
is sequential decoding \emph{essential}
for the superior~per- formance of PAC codes? 
Is it possible, or perhaps advantage\-ous, to replace the sequential decoder
in Figure\,\ref{fig:PAC_scheme} by~an~alter\-native decoding method?
We show that, indeed, similar~per\-formance can be achieved using list decoding,
provided the~list\linebreak
size $L$ is moderately large. This conclusion is 
illustrated~in~Fig\-ure\,\ref{fig:performance}, where we use a list 
of size $L = 128$ to closely match~the performance of the sequential
decoder.
It remains to be seen which of 
the two approaches is advantageous 
in terms~of~com\-plexity. While a comprehensive answer to this question
would require implementation in hardware, we carry out a qualitative
complexity comparison in \Sref{sec:list-vs-seq}. This comparison indicates that
list decoding has distinct advantages over sequential decoding in certain
scenarios. In particular, list decoding is certainly advantageous
in low-SNR regimes or in situations where the \emph{worst-case}
complexity/latency is the primary constraint.

\looseness=-1
\hspace*{-0.90ex}Another objective of this paper is to provide some insights~in-
to the remarkable performance of PAC codes. 
Although theo\-retical analysis of list decoding remains an open problem~even 
for conventional polar codes, it has~been observed in numerous
studies that list decoding 
quickly approaches the performance of \pagebreak
maximum-likelihood decoding with increasing list-size $L$.
As expected,
we find this to be the case for PAC codes as well 
(see Figure\,\ref{fig:ML_bound}). 
Fortunately, maximum-likelihood decoding of linear codes is reasonably
well understood: its performance~is governed by their weight distribution,
and can be well approximated by the union bound, especially at high SNRs.
Motivated by this observation, we use the method of \cite{LST12} to 
estimate~the number of low-weight codewords in PAC codes,~under~polar 
and RM rate profiles (introduced by \Arikan{}~\cite{Arikan19}). 
We find that PAC codes with the RM rate-profile
are superior to both polar codes (with or without CRC precoding) \emph{and}
the $(128,64,16)$ Reed-Muller code. For more on this, see Table\,\ref{table}
and Figure\,\ref{fig:union_bound}. 
These results suggest that the goal
of rate-profiling may be to optimize the weight distribution 
at low weights.

\vspace{1.08ex}
\subsection{Related Work}

\noindent
Numerous attempts have been made to improve the perform\-ance of polar codes 
at short block lengths.
Various approaches based on replacing 
successive-cancellation decoding with more advanced decoders
include 
list decoding~\cite{TV15}, 
sequential decoding~\cite{MT14}, and stack decoding~\cite{Niu12},
among others.
%
%
As 
shown later in this paper,
in Ar{\i}kan's PAC codes, convolutional precoding combined~with
rate-profiling can be regarded as replacing  
traditional frozen bits with \emph{dynamically frozen bits}.
Polar coding with {dynamically frozen bits} was first studied
by Trifonov and Miloslavska\-ya in~\cite{TM13}, although
the dynamic freezing patters in \cite{TM13}~and~\cite{Arikan19}
are very different.
Prior to~\Arikan's paper~\cite{Arikan19}, convolu\-tional 
precoding of polar codes was~propo\-sed in~\cite{FTY17} and
later studied in \cite{FVY19}.
Finally, the work of~\cite{RBV20}, which considers 
Fano and 
list decoding of PAC codes, is independent from and contemporaneous
with our results herein%
\footnote{The work of Rowshan, Burg, and Viterbo \cite{RBV20}, 
was posted on \texttt{arxiv.org} in February 2020, while our work
was submitted for review in January 2020.~~}\!.~~

\vspace{1.08ex}
\subsection{Paper Outline}

\noindent
\looseness=-1
The rest of this paper is organized as follows. 
We begin with~an overview on Ar{\i}kan's PAC codes in \Sref{sec:overview},
including both their encoding process and sequential decoding.
In \Sref{sec:list}, we present our list-decoding algorithm.
In \Sref{sec:list-vs-seq}, 
we compare it with sequential decoding, in terms of both 
performance and complexity.
In \Sref{sec:analysis}, we endeavor to acquire
some insight into the
remarkable performance of PAC codes.
First,~we show empirically that both sequential decoding and
list decoding thereof are extremely close to the ML decoding perform\-ance.
To get a handle on the latter,
we estimate the number of low-weight codewords in PAC codes (and polar codes) 
under different 
rate profiles. This makes it possible to approximate
the performance of ML decoding with a union bound.
We~conclude with a brief discussion in \Sref{sec:last}.

\vspace{2.70ex}
\section{Overview of Ar{\i}kan's PAC codes}
\label{sec:overview}
\vspace{0.54ex}

\noindent 
For details on conventional polar codes under
standard SC decoding, 
we refer the reader to Ar{\i}kan's seminal paper~\cite{Arikan09}.

Like polar codes, the block length $n$ of a~PAC 
code is also a power of $2$. That is, $n=2^m$ with $m\ge 1$. 
As shown in Figure\,\ref{fig:PAC_scheme}, 
the encoding process for an $(n,k)$ PAC code consists of\linebreak
the following three steps: \pagebreak
rate-profiling, convolutional precoding, and polar en\-coding. 
In the first step,
the $k$ data (information) bits~of~the data vector 
$\ddd$ are embedded into a data-carrier vector $\vvv$
of length $n$, at $k$ positions specified by 
an index set $\cA \subseteq \{0,1,\ldots,n\,{-}\,1\}$ with $|\cA|=k$.  
The remaining $n-k$~po\-sitions in $\vvv$ are frozen to zero.
\Arikan~\cite{Arikan19} used 
{\dfn rate-profiling} to refer to this step, along with the choice of
the index set~$\cA$.~~ 

\looseness=-1
Just like for polar codes, a careful choice of the index set $\cA$ 
is crucial to achieve good performance. 
\Arikan{} has proposed in \cite{Arikan19} 
two alternative approaches for selecting this set $\cA$.
The first approach, called \emph{\dfn polar rate-profiling},
proceeds as follows.\linebreak
Let $W_0,W_1,\dots,W_{n-1}$ be the $n$ bit-channels, defined
with respect to the conventional polar code of length $n$. 
In polar rate-profiling, $\cA$ is chosen so that
$\{W_i:i\in\!\cA\}$ consists of the $k$ best bit-channels in terms
of their capacity.
In other words,~the capacities of the $k$ bit-channels $\{W_i:i\in\!\cA\}$
are the $k$ highest\linebreak
values among $I(W_0),I(W_1),\ldots,I(W_{n-1})$.
The second app\-roach proposed in \cite{Arikan19} 
is called \emph{\dfn RM rate-profiling}. 
Let $\wt(i)$ denote the Hamming weight of the binary expansion of 
an~index $i$. In RM rate-profiling,~we simply pick the $k$ indices
of the highest weight, with ties resolved arbitrarily. 
In other words,
the set $\{\wt(i):i\in\!\cA\}$ consists~of the $k$ largest values 
among $\wt(0),\wt(1),\ldots,\wt(n-1)$.
Notably, \emph{without convolutional precoding}, this 
choice of $\cA$ generates Reed-Muller codes (as subcodes of
a rate-$1$ polar code).

\hspace*{-0.72ex}In 
the second step, the data-carrier vector $\vvv$ resulting from~the 
rate-profiling step is encoded using a rate-$1$ convolutional 
code generated by 
${\ccc} = (c_0,c_1,\ldots,c_{\nu})$, with $c_0 = c_\nu = 1$
(the latter can be assumed without loss of generality).
This produces another vector $\uuu = (u_0,u_1,\ldots,u_{n-1})$
of length $n$,
where\vspace{-0.36ex}
\begin{align*}
u_0 &=\, c_0v_0, \\
u_1 &=\, c_0v_1+c_1v_0, \\
u_2 &=\, c_0v_2+c_1v_1+c_2v_0,\quad 
\end{align*}
and so on. In general, every bit in ${\uuu}$ is a linear combination~of 
$(\nu+1)$ bits of $\vvv$ computed via the convolution operation:
\be{convolution}
u_i \,=\, \sum_{j=0}^{\nu} c_jv_{i-j}
\ee
where for $i-j < 0$, we set $v_{i-j} = 0$ by convention.
Alternati\-vely, this 
step can be viewed as a vector-matrix
multiplication $\uuu = \vvv{\bf T}$, 
where ${\bf T}$ is the upper-triangular Toeplitz matrix:
\be{Toepliz-def}
{\bf T} 
\ = \
\renewcommand{\arraycolsep}{1.08pt}
\begin{bmatrix}
\,c_0  & c_1    & c_2    & \cdots & c_\nu   & 0      & \cdots & 0    
\\[-1.26ex]
0      & c_0    & c_1    & c_2    & \cdots & c_\nu   &        & \vdots 
\\[-1.26ex]
0      & 0      & c_0    & c_1    & \ddots & \cdots & c_\nu  & \vdots 
\\[-1.26ex]
\vdots & 0      & \ddots & \ddots & \ddots & \ddots &       & \vdots 
\\[-1.26ex]
\vdots & \phantom{\ddots}& \ddots & \ddots & \ddots & \ddots & 0     & \vdots 
\\[-1.26ex]
\vdots &        &        & \ddots & 0      & c_0    & c_1   & c_2    
\\[-1.26ex]
\vdots &        &        &        & 0      & 0      & c_0   & c_1    
\\[-1.26ex]
0      & \cdots & \cdots & \cdots & \cdots & 0      & 0     & \,c_0\phantom{\vdots}    
\end{bmatrix}
\ee

In the third step, the vector $\uuu$ is 
finally encoded by a conventional polar encoder 
as the codeword $\xxx = \uuu{\bf P}_m$.
Here 
$$
{\bf P}_m 
\,=\,
{\bf B}_n
\renewcommand{\arraycolsep}{2.70pt}
\begin{bmatrix}
\,1 & 0\, \\
\,1 & 1\,
\end{bmatrix}^{\otimes m}
$$
where ${\bf B}_n$ is the $n\times n$ bit-reversal permutation matrix,
and~${\bf P}_m$ is known as the \emph{polar transform matrix}.

With reference to the foregoing discussion, the PAC code~in 
Figure\,\ref{fig:performance} is obtained via RM rate-profiling
using the rate-$1$ convolutional code generated by
${\ccc} = (1,0,1,1,0,1,1)$. This produces the 
$(128,64)$ PAC code of rate $\shalf$, which is the code studied 
by \Arikan{} in \cite{Arikan19}. 
This specific PAC code will serve as our primary running 
example throughout the paper.

On the decoding side, 
\Arikan{}~\cite{Arikan19} 
employs sequential decod\-ing of the underlying convolutional
code to decode the data-carrier vector $\vvv$.  
Under the frozen-bit constraints imposed~by rate-profiling, 
the rate-$1$ convolutional code 
becomes an irregular tree code. There are many different
variants of sequential decoding for irregular tree codes, 
varying in terms of both~the decoding metric used and the algorithm 
itself. \Arikan~\cite{Arikan19} uses the \emph{Fano sequential decoder},
described 
in~\cite{Fano63} and \cite{Gallager68}.
Notably, the path metrics at the input to the 
sequential 
decoder~are obtained via repeated 
calls to the successive-cancellation 
decoder for the underlying polar code,
as shown in Figure\,\ref{fig:PAC_scheme}.

\vspace{2.70ex}
\section{List Decoding of PAC Codes}
\label{sec:list}

\noindent 
\looseness=-1
One of our main objectives herein is to determine whether~seq\-uential
decoding of PAC codes (cf.~\Fref{fig:PAC_scheme}) can be replaced 
by list decoding. In this section, we show how list decoding of PAC
codes can be implemented efficiently. In the next section, we will consider
the performance and complexity of the result\-ing decoder, as compared
to the sequential decoder of \cite{Arikan19}.

\vspace{1.26ex}
\subsection{PAC Codes as Polar Codes with Dynamic Freezing}
\label{sec:dynamic}

\noindent 
To achieve efficient list decoding of PAC codes, we use~the~list-decoding
algorithm developed in~\cite{TV15}. The 
complexity of this algorithm
is $O(L n \log n)$, where $L$ is the list size. 
However, the algorithm of~\cite{TV15} decodes \emph{conventional
polar codes}. In order to make it possible to decode PAC codes with
(a modified version of) this algorithm, we first observe that PAC
codes can be regarded as \emph{polar codes with dynamically frozen bits}.

\looseness=-1
Polar coding with dynamically frozen bits was first introduced 
by 
Trifonov and Miloslavskaya in~\cite{TM13}, and later studied by the same
authors in \cite{TM16} and \cite{TT17}. Let us briefly describe~the 
general idea. In conventional polar coding, it is common practice
to set all frozen bits to zero. That is, $u_i = 0$ for all $i \in \cF$,
where $\cF \subset \{0,1,\ldots,n{-}1\}$ denotes the set of frozen 
indices. However, 
this choice is arbitrary:
we can set $u_i = 1$ for some $i \in \cF$ and $u_i = 0$ for other 
$i \in \cF$. \Arikan{} showed 
in 
\cite{Arikan09}
that on symmetric channels, this does not affect the performance.
What matters is that the frozen bits are {fixed} and, therefore, 
known \emph{a~priori} to the decoder.
In~\cite{TM13}, it was further observed that in order to be known
\emph{a~priori} to the decoder, the frozen bits do \emph{not} 
have to be fixed. Given $i \in \cF$, we can set
\be{dyn-enc}
u_i \,=\, f_i(u_0,u_1,\ldots,u_{i-1})
\ee
where $f_i$ is a fixed Boolean function (usually, a linear function)
known \emph{a~priori} to the decoder. For all $i \in \cF$, 
the decoder~can then decide as follows
\be{dyn-dec}
\widehat{u}_i \,=\, f_i(\widehat{u}_0,\widehat{u}_1,\ldots,\widehat{u}_{i-1})
\ee
where $\widehat{u}_0,\widehat{u}_1,\ldots,\widehat{u}_{i-1}$ are 
its earlier decisions. The encoding/ de\-cod\-ing
process in \eq{dyn-enc} and \eq{dyn-dec} 
is known as dynamic freezing.

\begin{algorithm}[t!]
\caption{List Decoder for PAC Codes}
\label{alg:listPAC}
	\KwIn{The received vector $\yyy$, the list size $L$, the~generator 
	${\ccc}=(c_0,c_1,\ldots,c_{\nu})$ 
        for the~convolutional precoder as global}
	\KwOut{Decoded codeword $\widehat{\xxx}$}
\BlankLine
	\tcp{Initialization}
	$\cdots$ lines 2--5 of Algorithm\,12 in \cite{TV15}\\
	{\color{blue}
	shiftRegisters $\leftarrow$ {\bf new} 2-D array of size $L\times(\nu+1)$ \\
	\For{$\ell=0,1,\ldots,L-1$}
	{
		shiftRegister[$\ell$] = $(0,0,\ldots,0)$
	}
	}
	\tcp{Main Loop}
	\For{$\varphi = 0,1,\ldots,n-1$} 
	{
		recursivelyCalcP($m,\varphi$) \\
		\eIf {$u_\varphi$ is frozen}
		{
			\For{$\ell = 0,1,\ldots,L-1$}
			{
				\If{activePath$[\ell]$ = {\bf false}}
				{\bf continue}
				{\color{blue}
				left-shift shiftRegister[$\ell$]
				by one, with~the~rightmost position 
				set to $0$ \\
				}
				$C_m\leftarrow$ getArrayPointerC($m,\ell$) \\
				{\color{blue}
				$(v_{\varphi-\nu},v_{\varphi-\nu+1},\ldots,v_\varphi)\leftarrow$ shiftRegister[$\ell$] \\
				\tcp{Set the frozen bit}
				$C_m[0][\varphi\mod 2]\leftarrow\sum_{j=0}^\nu c_jv_{\varphi-j}$ \\
				}
			}
		}
		{
			{\color{blue}
			continuePaths\_Unfzn($\varphi$)
			}
		}
		\If{$\varphi\mod 2=1$}
		{
			recursivelyUpdateC($m,\varphi$)
		}
	}
	\tcp{Get the best codeword in the list}
	$\cdots$ lines 17--24 of Algorithm\,12 in \cite{TV15}\\
	\Return $\widehat{\xxx} = (C_0[\beta][0])_{\beta=0}^{n-1}$
\end{algorithm}

\hspace*{-1ex}In 
order to explain how \Arikan's PAC codes~\cite{Arikan19}
fit into~the~dy\-namic freezing framework, let us first 
introduce some notation. With reference to 
\Sref{sec:overview}, for $i = 0,1,\ldots,n{-}1$,
let $\uuu_i$ and $\vvv_i$ denote the vectors
$(u_0,u_1,\ldots,u_{i})$ and $(v_0,v_1,\ldots,v_{i})$, respectively. 
Further, let ${\bf T}_{i,j}$ denote the submatrix of the Toe\-pliz
matrix ${\bf T}$ in \eq{Toepliz-def}, consisting of 
the first (topmost) $i+1$~rows and the first (leftmost) $j+1$ columns.
With this, it is easy to see
that 
$\uuu_i = \vvv_i \bT_{i,i}$ for all $i$.
The 
matrix $\bT_{i,i}$ is upper triangular with 
$\det\bT_{i,i} = c_0 = 1$. 
Hence it is invertible,~and~we~have 
$\vvv_i = \uuu_i \bT_{i,i}^{-1}$ for all $i$.
Now suppose that $i \in \cA^c$, so that $v_i$ is frozen to zero
in the rate-profiling step. Then we have
\be{PAC-freeze}
\uuu_i 
\,=\,
\vvv_{i-1} \bT_{i-1,i}
\,=\,
\bigl(\uuu_{i-1} \bT_{i-1,i-1}^{-1}\bigr) \bT_{i-1,i}
\ee
In particular, this means that 
the last bit $u_i$ of the vector $\uuu_i$ is 
an \emph{a priori} fixed linear
function of its first $i$ bits,~as~follows:
$$
u_i 
\,=\,
(u_0,u_1,\ldots,u_{i-1})\,
\bT_{i-1,i-1}^{-1} 
\bigl({0,\ldots,0},c_\nu,c_{\nu-1},\ldots,c_1\bigr)^t
$$
where $(0,\ldots,0,c_\nu,c_{\nu-1},\ldots,c_1)^t$ represents 
the last column~of the matrix $\bT_{i-1,i}$.
Clearly, the above is a special case of dynamic freezing 
in \eq{dyn-enc}.
Moreover, it follows that the set $\cF$ of indices that are dynamically frozen
is precisely the same~as~in the rate-profiling step, that is $\cF = \cA^c$.

If $i\,{\in}\,\cA$, then $v_i$ is an information bit,
but the value of $u_i$~is determined not only 
by $v_i$ but by 
$v_{i-1},v_{i-2},\ldots,v_{i-\nu}$ as well.
Thus when representing PAC codes as polar codes, the information bits
may be also regarded as dynamic.

Finally, note that in implementing the PAC decoder, there~is no need
to actually invert a matrix as in \eq{PAC-freeze}. Instead, 
we~succes\-sively compute the vector
$
\widehat{\vvv} = (\widehat{v}_0,\widehat{v}_1,\ldots,\widehat{v}_{n-1})
$
as follows.~\,If $i \,{\in}\, \cA^c$, set $\widehat{v}_i = 0$. Otherwise, set
\be{de-convolution}
\widehat{v}_i ~=~ \widehat{u}_i \,-\, \sum_{j=1}^{\nu} c_j\widehat{v}_{i-j}
\ee
where the value of $\widehat{u}_i$ is provided by the polar decoder.
Given
$\widehat{v}_{i},\widehat{v}_{i-1},\ldots,\widehat{v}_{i-\nu}$,
the values of the dynamically frozen bits $\widehat{u}_i$ 
for $i \,{\in}\, \cA^c$ can be computed using \eq{convolution}.
This computation, along with the one in \eq{de-convolution}, 
takes linear time. All that is required~is~additional memory
to store the vector
$
\widehat{\vvv} = (\widehat{v}_0,\widehat{v}_1,\ldots,\widehat{v}_{n-1})
$.

\begin{algorithm}[t]
\caption{continuePaths\_Unfzn (PAC\,version)}
\label{alg:contUnfzn}
	\KwIn{phase $\varphi$}
	\BlankLine
	$\cdots$ lines 1--18 of Algorithm\,13 in \cite{TV15}\\
	\tcp{Continue relevant paths}
	\For{$\ell = 0,1,\ldots,L-1$} 
	{
		\If{contForks$[\ell][0]$ = {\bf false} and 
			\hspace*{5.4ex}\mbox{contForks$[\ell][1]$ = {\bf false}}}
		{\bf continue}
		$C_m\leftarrow$ getArrayPointer\_C$(m,\ell)$ \\
		{\color{blue}
		left-shift shiftRegister[$\ell$]
		by one, with~the~rightmost position 
		set to $0$ \\
		$(v_{\varphi-\nu},v_{\varphi-\nu+1},\ldots,v_\varphi)\leftarrow$ 
                       shiftRegister[$\ell$] \\
		}
		\eIf{contForks$[\ell][0]$ = {\bf true} and 
			contForks$[\ell][1]$ = {\bf true}}
		{
			{\color{blue}
			$C_m[0][\varphi\mod 2] \leftarrow \sum_{j=0}^\nu c_jv_{\varphi-j}$ \\
			}
			$\ell'\leftarrow$ clonePath$(\ell)$ \\
			{\color{blue}
			shiftRegister[$\ell'$] $\leftarrow$ shiftRegister[$\ell$] \\
			flip the rightmost bit of shiftRegister[$\ell'$] \\
			}
			$C_m\leftarrow$ getArrayPointer\_C$(m,\ell')$ \\
			{\color{blue}
			$(v'_{\varphi-\nu},v_{\varphi-\nu+1},\ldots,v'_\varphi)\leftarrow$ shiftRegister[$\ell'$] \\
			$C_m[0][\varphi\mod 2] \leftarrow \sum_{j=0}^\nu c_jv'_{\varphi-j}$ \\
			}
		}
		{
			\eIf{contForks$[\ell][0]$ = {\bf true}}
			{
				{\color{blue}
				\If{$\sum_{j=0}^\nu c_jv_{\varphi-j} = 1$}
				{
					flip the rightmost bit of shiftRegister$[\ell]$ 
				}
				}
				set $C_m[0][\varphi\mod 2]\leftarrow 0$
			}
			{
				{\color{blue}
				\If{$\sum_{j=0}^\nu c_jv_{\varphi-j} = 0$}
				{
					flip the rightmost bit of shiftRegister$[\ell]$ 
				}
				}
				set $C_m[0][\varphi\mod 2]\leftarrow 1$
			}
		}
	}
\end{algorithm}

\vspace{1.26ex}
\subsection{List Decoding of PAC codes}

\noindent 
Representing PAC codes as polar codes with dynamically frozen bits
makes it possible to adapt existing algorithms for
successive-cancellation list decoding of polar codes to 
decode PAC codes. 

There are, however, several important differences.
For exam\-ple, for conventional polar codes, whenever the list 
decoder~encounters a frozen index $i \,{\in}\, \cF$, all the paths in the
list-decoding tree are extended in the same way, by setting
$\widehat{u}_i = 0$. 
For PAC codes, since the freezing is dynamic, different paths 
are~potent\-ially extended differently, depending upon the previous
decisi\-ons along the path.

\hspace*{-0.54ex}In general, our list decoder for PAC codes maintains~the~same
data structure as the successive-cancellation list decoder
in~\cite{TV15}. In addition, for a list of size $L$, we introduce $L$
auxiliary shift registers --- one for each path. Each shift register 
stores~the~last $\nu$ bits of the vector
$
\widehat{\vvv} = (\widehat{v}_0,\widehat{v}_1,\ldots,\widehat{v}_{n-1})
$,
computed as~in~\eq{de-convolution}, for the corresponding path.

Algorithm\,\ref{alg:listPAC} and Algorithm\,\ref{alg:contUnfzn} 
provide the full details of our list decoding algorithm for PAC codes. 
These algorithms
fit~in\-to the same general mold 
as Algorithms 12 and 13~of~\cite{TV15},~with 
the differences highlighted in blue.

\vspace{2.34ex}
\section{List Decoding versus Sequential Decoding}
\vspace{0.36ex}
\label{sec:list-vs-seq}

\noindent 
\looseness=-1
We now compare list decoding 
of PAC codes with sequential\linebreak
decoding,
in terms of both performance and complexity. For list decoding, we
use the algorithm 
of \Sref{sec:list}.
For sequential
decoding, we employ exactly the same Fano decoder that was used by
\Arikan{} in \cite{Arikan19}. We are grateful to Erdal \Arikan~for
sharing the details of his decoding algorithm.
We do not disclose these details here, instead referring the
reader to \cite{Arikan19}.

Our main conclusion is that sequential decoding is \emph{not}
essential in order to achieve the remarkable performance of PAC codes:
similar performance can be obtained with list decoding, providing
the list size is sufficiently 
large.
As far as complexity, sequential decoding is generally better at 
high SNRs and in terms of average complexity, while list decoding
is advantageous in terms of worst-case complexity and at low SNRs.~~

\vspace{0.90ex}
\subsection{Performance Comparison}
\label{sec:comp-performance}

\begin{figure}[t!]
\centering
\includegraphics[trim=60 220 60 220, width=3.24in]%
{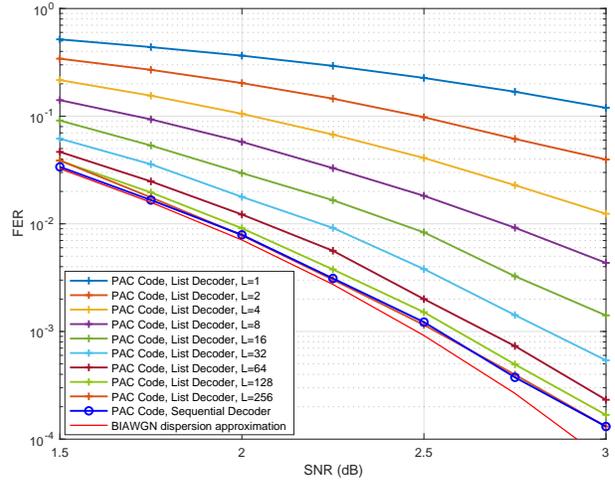}
\vspace{2.70ex}
\caption{Performance of PAC codes under list decoding}
\vspace{-1.98ex}
\label{fig:list_performance}
\end{figure}

\noindent 
\looseness=-1
Figure\,\ref{fig:list_performance} summarizes simulation results 
comparing the perform\-ance of \Arikan's Fano decoder from~\cite{Arikan19}
with our list deco\-ding algorithm, as a function of the list size $L$. 
The underlying PAC code is the same as in \Fref{fig:performance}; it is
the $(128,64)$ PAC~code obtained via RM rate-profiling 
(see \Sref{sec:overview}).
The underlying channel is the binary-input additive white Gaussian~noise
(BIAWGN) channel.

\looseness=-1
As expected, the performance of list decoding steadily improves with
increasing list size until we reach a point of dimin\-ishing returns.
For $L = 128$, the list-decoding performance~is
very close to that of sequential
decoding, while for $L = 256$~the two curves virtually 
coincide over the entire range of SNRs.~~

It should be pointed out that the frame error rate (FER) reported 
for sequential decoding in 
Figures \ref{fig:performance} and \ref{fig:list_performance}
is due~to~two different mechanisms of error/failure.
In some cases, the sequential decoder reaches the end of the search
tree (see \Fref{fig:tree}) producing an incorrect codeword. These
are decoding errors. 
In other cases, the end of the search tree is never reached; instead,
the computation is aborted once it exceeds a predetermined cap on the
number of cycles. These are decoding~fail-\linebreak ures. 
As in \cite{Arikan19}, the FER plotted in \Fref{fig:list_performance}
counts all~the~cases wherein the transmitted codeword is not produced
by the decoder: thus it is the \emph{sum of the error rate and the
failure rate}. The table below shows what fraction of such cases were
due~to decoding failures:\vspace{0.54ex}
\begin{center}
\small\footnotesize
\begin{tabular}{|l|@{~\,}c@{~\,}|@{~\,}c@{~\,}|@{~\,}c@{~\,}|@{~\,}c@{~\,}|@{~\,}c@{~\,}|@{~\,}c@{~\,}|}
\hline
\rule{0cm}{2.10ex} \hspace*{-3mm} SNR [dB] & 
 1.00~ &  1.25~ &  1.50~ &  1.75~ &  2.00~ &  2.25~ \\
\hline
\rule{0cm}{2.10ex} \hspace*{-3mm} \% of failures & 
4.53\% & 3.56\% & 1.86\% & 1.38\% & 1.01\% & 0.29\% \\
\hline
\end{tabular}
\vspace{0.90ex}
\end{center}
A decoding failure was declared in our simulations
whenever the number of cycles (loosely speaking, 
cycles count for\-ward and backward movements along the search tree in
the Fano decoder) exceeded $1,300,000$.
This is exactly the same cap on the number of cycles 
that was used by \Arikan\ in \cite{Arikan19}.
Overall, the foregoing table 
indicates that increasing 
this cap 
would {not} improve the performance significantly.

The FER for list decoding is also due to two distinct error
mechanisms. In some cases, the transmitted codeword is not
among the $L$ codewords generated by our decoding algorithm.
In other cases, it \emph{is} on the list of codewords generated,
but~it~is not the most likely among them. Since the list decoder
selects the most likely codeword on the list as its ultimate 
output, this leads to a decoding error. We refer to such
instances~as~select\-ion errors.
The table below shows the fraction of selection~er\-rors
for lists of various sizes:\vspace{0.72ex}
\begin{center}
\small\footnotesize
\begin{tabular}{|l|@{~\,}c@{~\,}|@{~\,}c@{~\,}|@{~\,}c@{~\,}|@{~\,}c@{~\,}|@{~\,}c@{~\,}|@{~\,}c@{~\,}|@{~\,}c@{~\,}|}
\hline
\rule{0cm}{2.10ex} \hspace*{-3mm} SNR\,[dB]\hspace*{-0.72ex} & 
 1.50~ &  1.75~ &  2.00~ &  2.25~ &  2.50~ &  2.75~ &  3.00\\
\hline
\rule{0cm}{2.10ex} \hspace*{-3mm} $L = 64$ & 
32.1\% & 32.2\% & 32.5\% & 32.3\% & 29.4\% & 36.7\% & 39.6\% \\
\hline
\rule{0cm}{2.10ex} \hspace*{-3mm} $L = 128$ & 
50.0\% & 51.6\% & 54.6\% & 53.6\% & 58.4\% & 60.4\% & 63.2\% \\
\hline
\rule{0cm}{2.10ex} \hspace*{-3mm} $L = 256$ & 
66.2\% & 71.0\% & 75.2\% & 78.0\% & 79.9\% & 83.6\% & 82.8\% \\
\hline
\end{tabular}
\vspace{0.90ex}
\end{center}
This indicates that the performance of list decoding would~further
improve (at least, for $L \,{\ge}\,\, 64$) if we could somehow~increase
the minimum distance of the underlying code, 
or other\-wise aid the decoder in selecting from the list
(e.g.~with CRC).~ 

Finally, we also include in 
Figures \ref{fig:performance} and \ref{fig:list_performance}
the BIAWGN~dis\-persion-bound approximation for binary codes of rate
$\shalf$ and length $128$. The specific curve plotted in
Figures \ref{fig:performance} and \ref{fig:list_performance}~is~the 
so-called \emph{normal approximation} of the 
dispersion bound~of~Po\-l\-yanskiy, Poor, and Verdu~\cite{PPV10}.
Our curve coincides with those given in \cite[Figure\,1]{Durisi+19}
and \cite[Figure\,6]{GB19}. 
Note that a more accurate bound can 
be derived using the methods of Erseghe~\cite{Erseghe16}, but this
is not critical for our purposes.
It is clear from \pagebreak
Figures \ref{fig:performance} and \ref{fig:list_performance}
that the performance of the $(128,64)$ PAC code, under both 
sequential decoding and list decoding with $L \ge 128$,~is
close to the best achievable performance. 

\vspace{1.26ex}
\subsection{Complexity Comparison}
\label{sec:comp-complexity}

\noindent 
A comprehensive complexity analysis of list decoding versus sequential 
decoding of PAC codes in practical applications~is
likely to require 
algorithmic optimization and implementation in hardware. 
In the case of 
list decoding, this should be relat\-ively easy 
based upon 
our representation of PAC codes as polar codes with dynamically
frozen bits (see \Sref{sec:dynamic})~in~conjunction with 
existing work on efficient hardware
implementation of polar list decoders (see \cite{SGVTG14,SGVTG16}, 
for example).
On the other hand, we are not aware of any existing implementations
of sequential decoding in hardware. Such implementation may be
challenging due to variable running time, which depends~on the
channel noise, and complex control logic~\cite{Balatsoukas}.

\begin{figure}[t!]
  \centering
\includegraphics[width=3.00in]{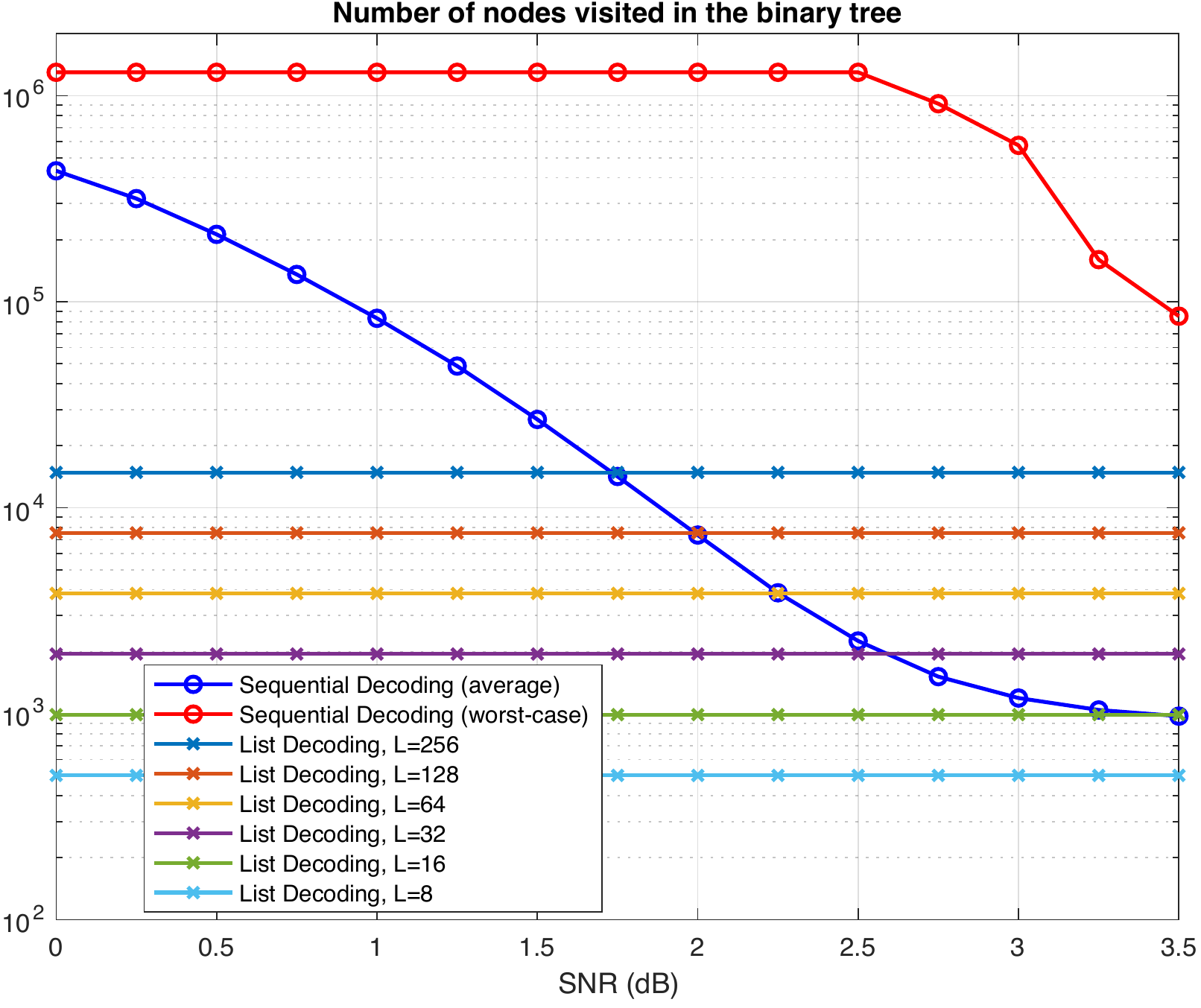}
\vspace{0.36ex}
\caption{Sequential decoding vs.\ list decoding complexity~compari\-son: 
Number of nodes visited in the polar search tree per codeword~~}
\vspace{1.44ex}
  \label{fig:complexity_node}
\end{figure}

\begin{figure}[t!]
  \centering
  \includegraphics[width=3.00in]{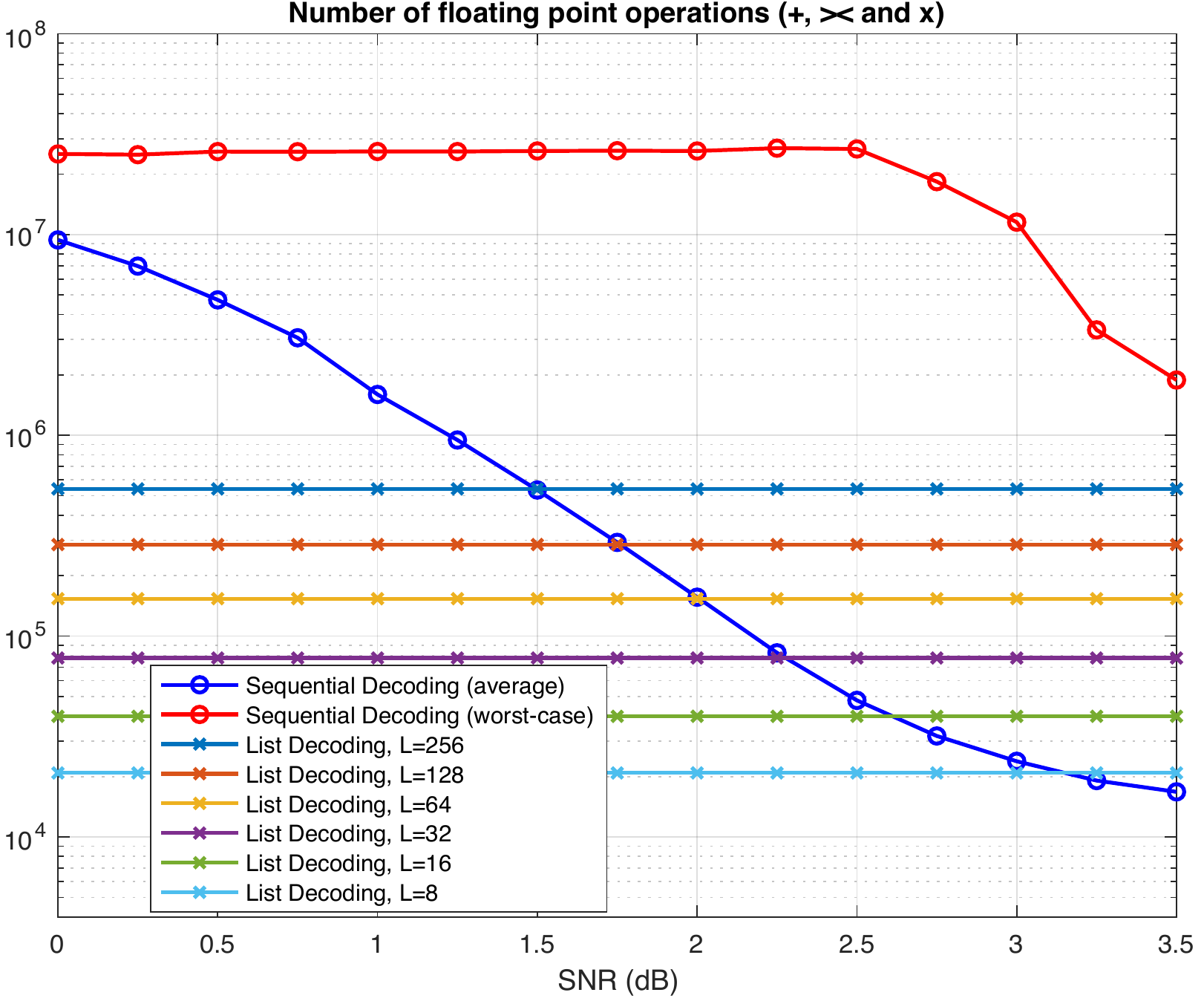}
\vspace{0.36ex}
\caption{Sequential decoding vs.\ list decoding complexity~compari\-son: 
Number of floating-point operations per decoded codeword}
  \label{fig:complexity_f}
\end{figure}

\looseness=-1
In this section, we provide a qualitative
comparison of list
decoding versus sequential decoding 
using two generic~complexity 
metrics: the number of nodes visited in the polar
search tree and the total number of floating-point operations~perfor\-med 
by the decoder. The results we obtain for the two
metrics, summarized in 
Figures \ref{fig:complexity_node} and \ref{fig:complexity_f},
are consistent with each other.~

\begin{figure}[t!] 
  \centering
\includegraphics[width=2.70in]{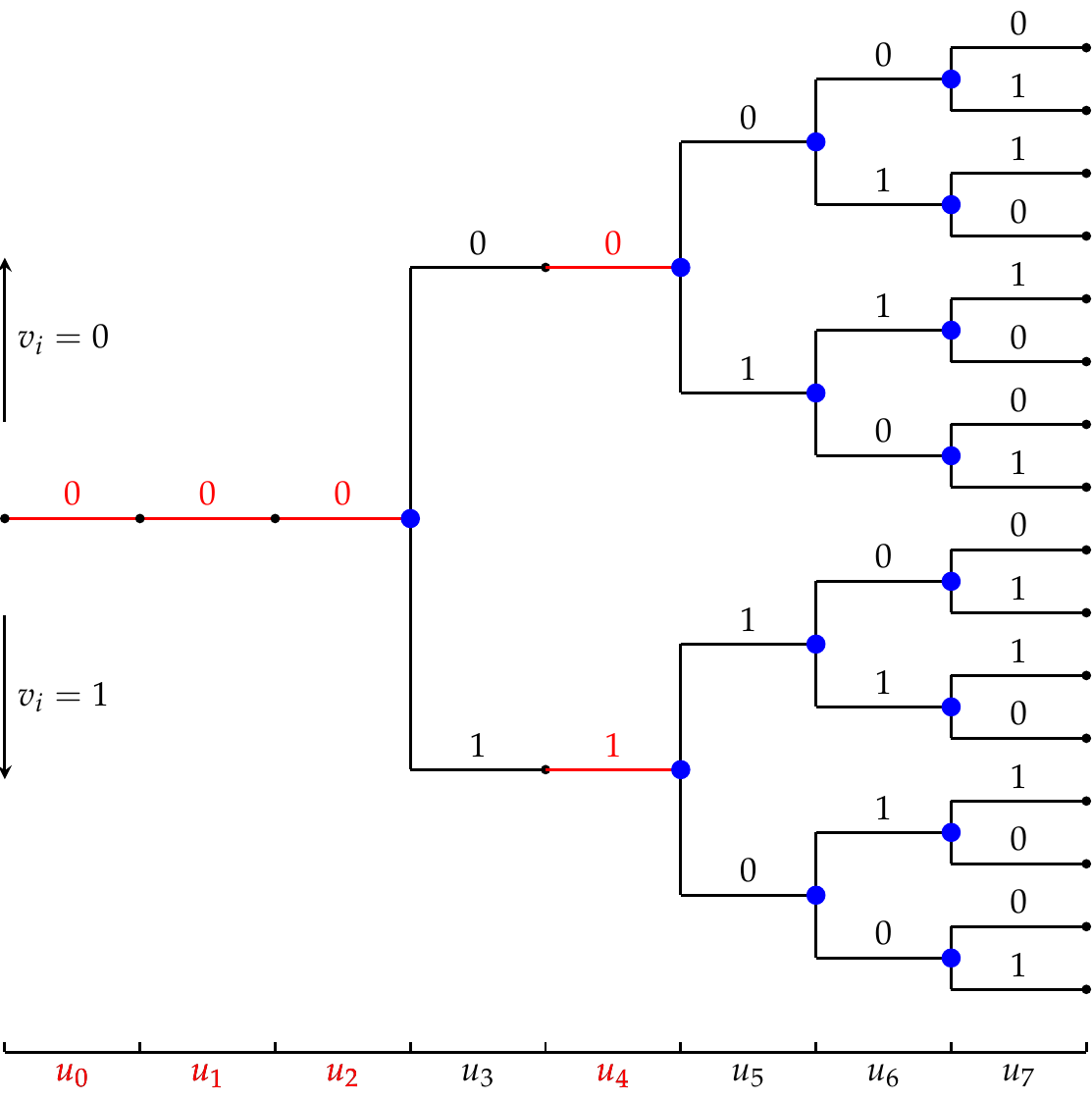}
\vspace{0.45ex}
\caption{An example of the polar search tree, reproduced from~\cite{Arikan19}}
  \label{fig:tree}
\end{figure}

The polar search tree, shown schematically in \Fref{fig:tree}, represents
all possible inputs $\uuu = (u_0,u_1,\ldots,u_{n-1})$ to the polar encoder.
It is an irregular tree with $n+1$ levels containing $2^k$ paths.
If $i\kern-0.5pt\in\kern0.5pt\!\cA^c$ then all nodes at level $i$ 
have a single outgoing edge, as $u_i$ is dynamically frozen in this case.
In contrast~with conventional polar codes, 
these edges may be labeled 
differently (cf.~$u_4$ in \Fref{fig:tree}).
If $i\kern-0.5pt\in\!\cA$ then all nodes at level $i$ have two 
outgoing edges.
In this framework, both list decoding and sequential decoding can be regarded 
as tree-search~algorithms that try to identify the most likely path in the tree.
The list decoder does so by following $L$ paths in the tree, 
from the root~to\linebreak
the leaves, and selecting the most likely one at the end.
The Fano sequential decoder 
follows only one path, 
but has many back-and-forth movements during the decoding process.

\hspace*{-0.54ex}For the sake of qualitative comparison, we take
the total~num\-ber of nodes the two algorithms visit 
in the tree as one reasonable proxy of their complexity.
In doing so, we disregard~the nodes at the frozen levels,
and count only those nodes 
that~have two outgoing edges 
(colored blue in \Fref{fig:tree});
we call them~the \emph{decision nodes}.
\Fref{fig:complexity_node} shows the number of decision~nodes 
visited by the two decoding algorithms as a function of SNR.~~~ 

For sequential decoding, two phenomena are immediately apparent
from \Fref{fig:complexity_node}. First, there is a tremendous gap
bet\-ween worst-case complexity and average complexity. For most SNRs,
the worst-case complexity is dominated by decoding~fai\-lures, which
trigger a computational timeout upon reaching~the cap on the
number of cycles (see \Sref{sec:comp-performance}). 
Clearly, reduc\-ing this cap would also reduce the worst-case
complexity.~On the other hand, for SNRs higher than $2.50$\,dB, 
de\-coding failures were not observed. Thus, beyond $2.50$\,dB, 
the worst-case complexity gradually decreases, as expected. 
Another~phenom\-enon apparent from \Fref{fig:complexity_node} is
that the average complexity is highly dependent on SNR.
This is natural since the processing in the Fano sequential decoder
depends on the channel noise.\linebreak The less noise there is, 
the less likely is the sequential decoder to roll back in its search 
for a better path.

Neither of the two phenom\-ena above is present for list decoding:
the worst-case complexity is {equal} to the average complexity, and
both are unaffected by SNR. 
The resulting curves in Figures
\ref{fig:complexity_node} and \ref{fig:complexity_f} are flat,
since the complexity of list decoding depends only on the list 
size $L$ and the code dimension~$k$.~

In fact, the number of decision nodes visited by the list~dec\-o\-der
in the polar search tree can be easily computed as follows.\linebreak
First assume, for simplicity, that $L$ is a power of $2$.
As the~list decoder proceeds from the root to the leaves, 
the number of paths it traces doubles for every
$i\kern-0.5pt\in\!\cA$
until it reaches $L$.
The number of decision nodes 
it visits during this process is given by 
$
1 + 2 + 4 + \cdots + L = 2L-1
$.
After reaching $L$ paths, the decoder visits $L$ decision nodes 
at every one of the remaining 
$k-\log_2L$ levels that are not frozen.
Thus the total number~of decision nodes visited 
is 
$
L(k+2-\log_2 L) - 1
= O(kL)
$.
If $L$ is not a power of $2$, 
this counting argument
readily general\-izes, and the number of decision nodes visited is
given~by~~~~
\be{node-count}
L\bigl(k+1 - \lceil \log_2 L\rceil\bigr) +\, 2^{\lceil \log_2 L\rceil} - 1
\ = \
O(kL)
\ee

As another 
metric of 
complexity of the two algorithms, we count
the total number of additions, comparisons, and multipli\-cations 
of floating-point numbers throughout~the decoding process. The
results of this comparison are compiled in \Fref{fig:complexity_f}.
The number of floating-point operations is a~more precise measure
of complexity than the number of decision 
nodes visited in
the search tree. Yet we observe exactly the 
same pattern 
as in \Fref{fig:complexity_node}.
For list decoding, it is no longer possible to give 
a simple expression as in \eq{node-count}, but the 
complexity is still inde\-pendent of SNR, resulting in 
flat curves. For sequential decod\-ing, we 
observe the same two phenomena discussed earlier 
in connection with \Fref{fig:complexity_node}.
In particular, the worst-case~comple\-xity remains
prohibitive even at high SNRs. 

In summary, our qualitative comparison suggests that,
for a similar level of performance, sequential decoding
is clearly advantageous in terms of average-case complexity
at high SNRs. However, list decoding may have 
distinct advantages in low-SNR regimes or in situations 
where the worst-case complexity/latency is the primary constraint.

\vspace{2.52ex}
\section{Performance Analysis for PAC codes}
\vspace{0.36ex}
\label{sec:analysis}

\noindent 
\looseness=-1
In this section, we study the performance of PAC codes under\linebreak 
the assumption of maximum-likelihood (ML) decoding. To~this end,
we estimate computationally the number of low-weight codewords 
in PAC codes (and other codes), then 
combine~these estimates with the union bound. 
First, 
we explain why analysis of performance under 
ML~decoding makes sense in our setting.

\vspace{0.72ex}
\subsection{Sequential Decoding versus ML Decoding}
\label{sec:ML}

\noindent
It has been observed~in~several papers that for polar codes,
list decoding rapidly approach\-es the performance of ML
decoding with increasing list-size~$L$.
In this section, as expected, 
we find this to be the case for \Arikan's $(128,64)$
PAC code as well.

\begin{figure}[t!]
\centering
\includegraphics[trim=60 220 60 220, width=2.97in]{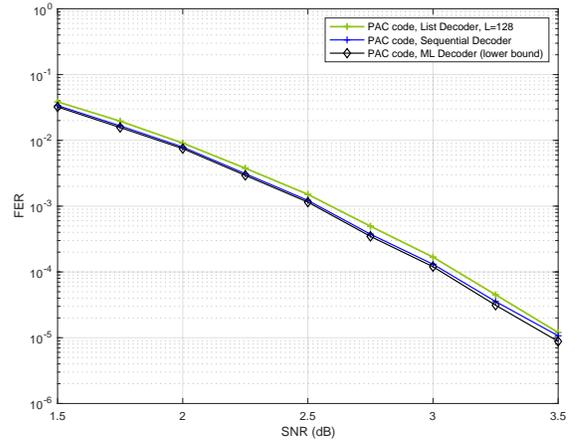}
\vspace{2.70ex}
\caption{Performance of the PAC code under ML decoding}\vspace*{-0.90ex}
\label{fig:ML_bound}
\end{figure}

\looseness=-1
\Fref{fig:ML_bound} shows a bound on the frame error-rate of ML decoding
obtained in our simulations. \,This is a \emph{lower bound},~in the sense that 
the actual simulated performance of ML decoding could be only worse --- even
closer to the other two curves 
(for sequential decoding and list decoding)
shown in \Fref{fig:ML_bound}.
The bound was generated using the Fano sequential decoder,~as\linebreak 
follows.
Every time the Fano decoder makes an error, 
we compare the likelihoods of the transmitted path 
and the path produced by the decoder.
If the decoded path has a higher path-metric
(likelihood), then the ML decoder will surely 
make~an error in this instance as well.
We count such instances~to~generate the lower bound.
This method of estimating ML performance in simulations
is very similar to the one introduced in \cite{TV15}
for polar codes, except that \cite{TV15} used list
decoding. 

\hspace*{-0.90ex}%
\Fref{fig:ML_bound} provides strong evidence that it makes
sense~to~study PAC codes under ML decoding in order to 
gain insights into their performance under sequential decoding,
since the two~are remarkably close.
\Fref{fig:ML_bound} also reveals 
one of the reasons~why \Arikan's PAC codes are so good
at short blocklengths: they~can be efficiently decoded
with near-ML fidelity. 

\vspace{1.08ex}
\subsection{Weight Distributions and Union Bounds}
\label{sec:weight-distributions}

\noindent
\hspace*{-0.36ex}We now study the weight distribution of the 
$(128,64)$ PAC~code in order to develop 
analytical understanding of its performance under ML decoding.
Specifically, we use the method of~\cite{LST12}~to estimate 
the number of low-weight codewords in this code.~

\begin{figure}[t!]
\centering
  \includegraphics[width=3.00in]{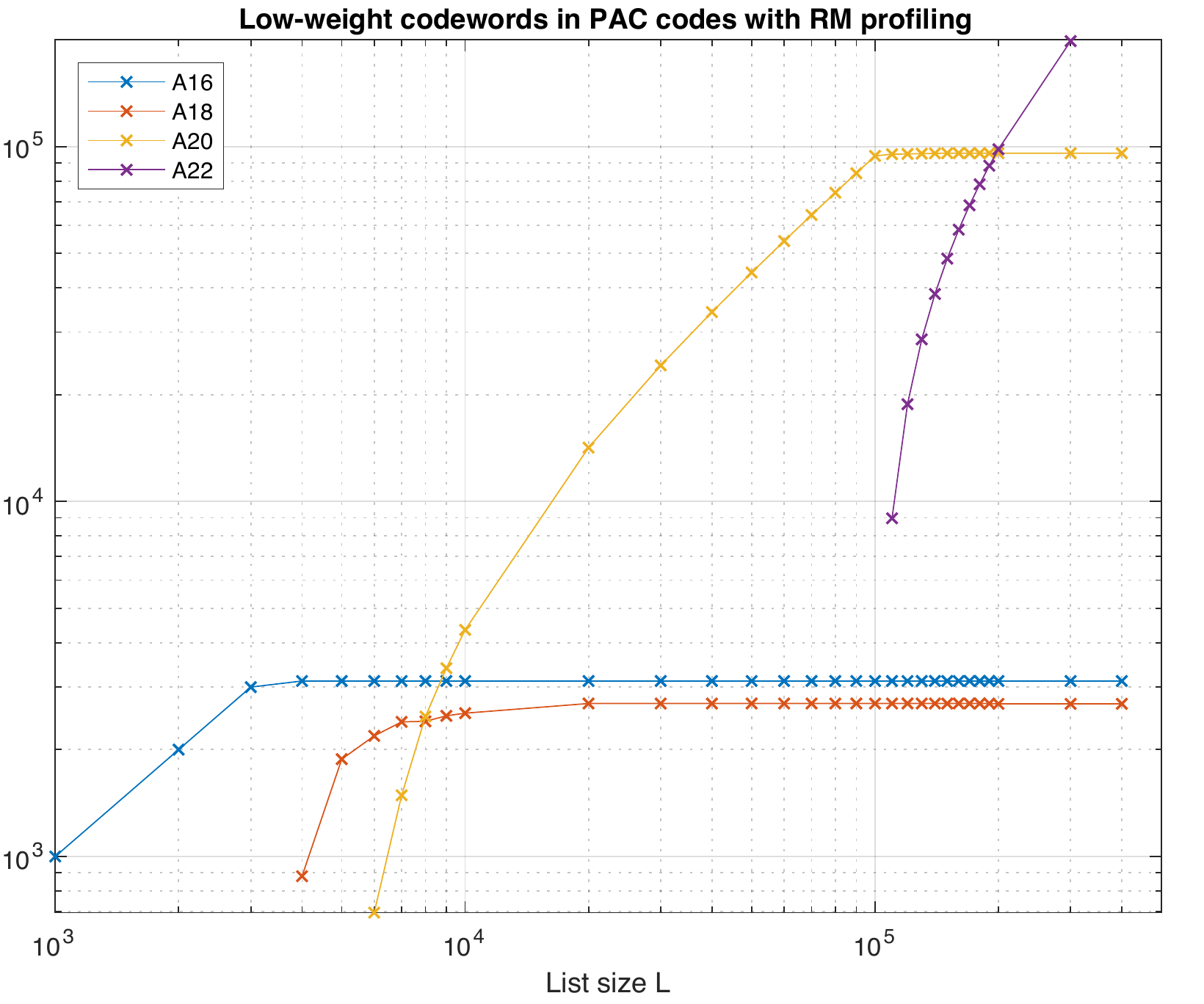}
\vspace{-.00ex}
\caption{Low-weight codewords in the $(128,64)$ PAC code} 
  \label{fig:weight_distribution}
\vspace{0.00ex}
\end{figure}

Consider the following experiment devised in \cite{LST12}.
Transmit the all-zero codeword in the extremely high SNR regime,
and use list decoding to decode the channel output. It is reasonable
to expect that in this situation, the list decoder will produce
codewords of low weight. As $L$ increases, 
since the decoder is forced
to generate a list of size exactly $L$, more and more low-weight 
codewords emerge. The results of this experiment for the
$(128,64)$ PAC code are shown in \Fref{fig:weight_distribution}
as a~function of the list size.
We can see that the only weights observed for $L$ up to $400,000$
are $16,18,20,$ and $22$. Moreover, 
$A_{16} \ge 3120$,\linebreak 
$A_{18} \ge 2696$, 
and
$A_{20} \ge 95828$
(cf.~Table\,\ref{table}).
These numbers\linebreak
are \emph{lower bounds} on the weight
distribution of the code. However, the fact that the
curves in \Fref{fig:weight_distribution} \emph{saturate}
at these values provides strong evidence that these bounds
are exact, and that codewords 
of other low weights do not exist.\pagebreak[3.99]

\looseness=-1
For comparison, we have 
used the same method to estimate\linebreak the number of 
low-weight codewords in other relevant codes of rate $\shalf$\kern0.75pt, 
including polar codes 
(with and without CRC precoding), the self-dual Reed-Muller code,
and the PAC code with po\-lar rate-profile.
The results are compiled in Table\,\ref{table} 
below.\vspace{0.90ex}

\begin{table}[h!]
\centering
\begin{tabular}{|l|c|c|c|c|c|c|}
\hline
 \rule{0cm}{2.10ex} \hspace*{-3mm}                 & 
$A_8$ & $A_{12}$ & $A_{16}$ & $A_{18}$ & $A_{20}$ & $A_{22}$ 
\\\hline
 \rule{0cm}{2.10ex} \hspace*{-3mm} Polar code            & 
4     & 0        & 68856    & 0       & $>10^5$ & -   
\\\hline
 \rule{0cm}{2.10ex} \hspace*{-3mm} Polar code, CRC8\hspace{-0mm}  
& 20  & 173      & \hspace{-1mm}$\ge7069$\hspace{-1mm}  & -  & -  & -          
\\\hline
      \rule{0cm}{2.10ex} \hspace*{-3mm} Reed-Muller            & 0     & 0        & 94488    & 0        & 0        & 0          \\\hline
      \rule{0cm}{2.10ex} \hspace*{-3mm} PAC, polar profile          & 48    & 0        & 11032    & 6024     & \hspace{-1mm}$>10^5$\hspace{-1mm}  & -                \\\hline
      \rule{0cm}{2.10ex} \hspace*{-3mm} PAC, RM profile          & 0     & 0        & 3120     & 2696     & 95828    & \hspace{-1mm}$>10^5$\hspace{-1mm}          \\\hline
\end{tabular}
\vspace{1mm}
\caption{Number of low-weight codewords in certain relevant codes}
\label{table}
\vspace{0.18ex}
\end{table}

\looseness=-1
Again, 
the numbers~in Table\,\ref{table} should be
regarded as lower bounds, which we conject\-ure to be exact
(except for the Reed-Muller code whose weight distribution 
is known~\cite{STK71}). Assuming 
this conjecture, we expect the performance of 
the $(128,64)$ PAC code under ML decoding to be superior 
to all other polar and PAC codes in the table, since its 
minimum distance is twice as high. 
Interestingly, this
code~is also superior~to~the self-dual Reed-Muller 
code. The two codes have the~same~min\-imum distance, but the
PAC code has significantly 
less code\-words at this
distance (by a factor of about $30$).

\begin{figure}[t!]
\centering
$\,$\\[-0.36ex]
  \includegraphics[width=3.18in]{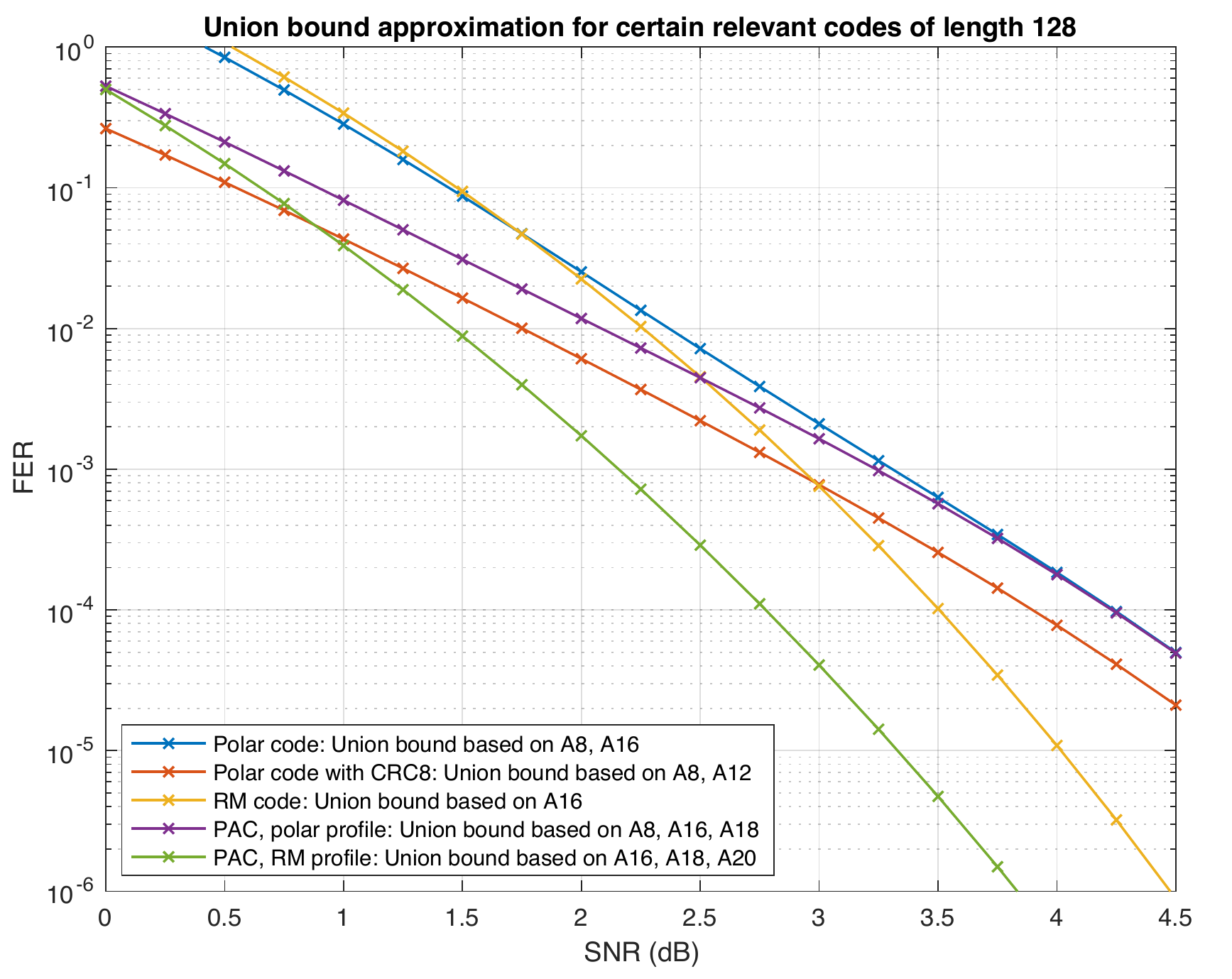}
\vspace{-0.18ex}
\caption{Truncated union bound for certain codes of length $128$}
  \label{fig:union_bound}
\end{figure}

\begin{figure}[t!]
\centering
$\,$\\[2.70ex]
  \includegraphics[width=3.18in]{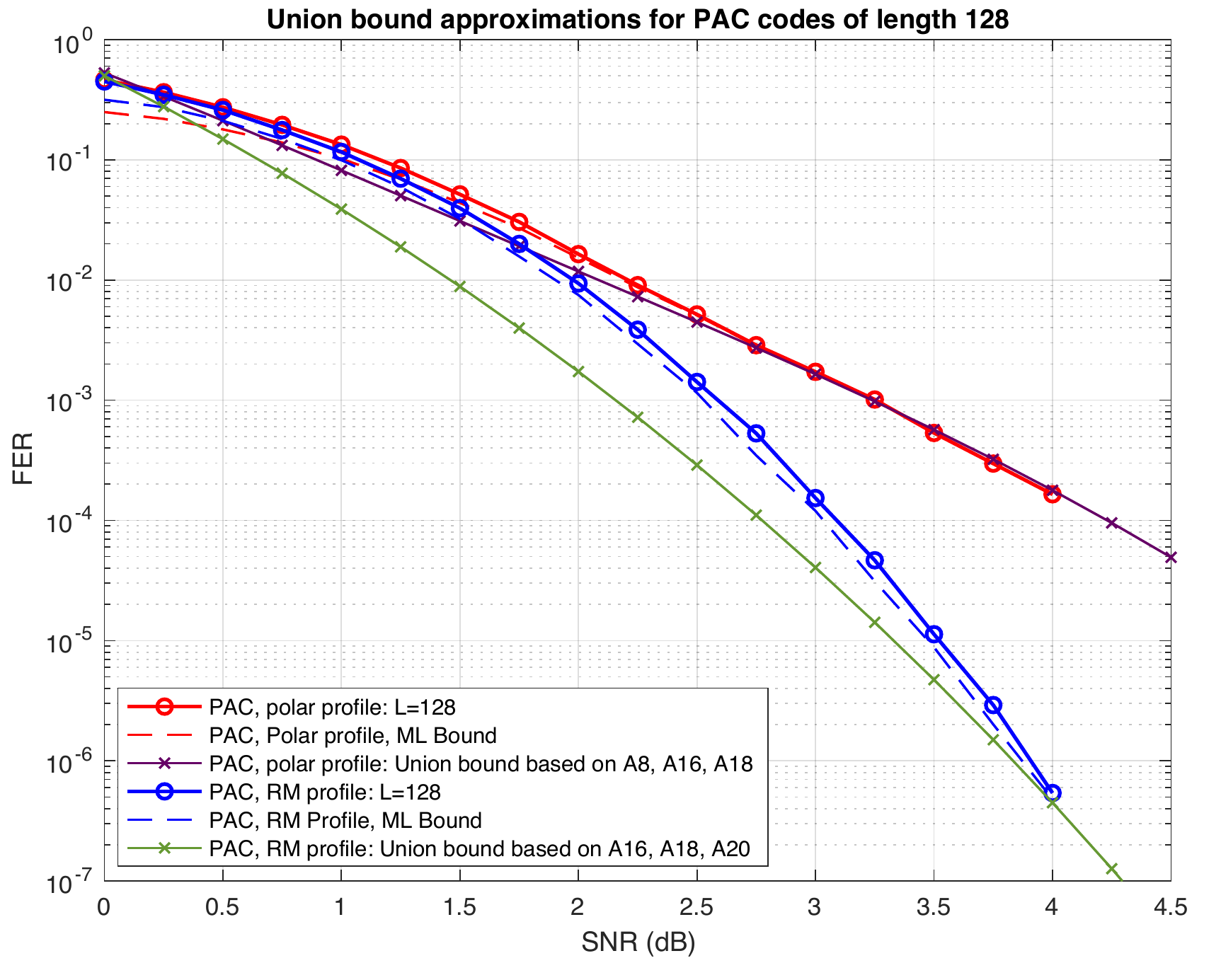}
\vspace{-0.36ex}
\caption{Truncated union bound vs.\ performance for two PAC codes}
  \label{fig:union_boundPAC}
\end{figure}

These observations are corroborated in \Fref{fig:union_bound},
where we plot the truncated union bound based on the partial 
weight distributions compiled in Table\,\ref{table} (with
all other terms set to zero).\linebreak 
It is well known that the performance of a linear code 
under ML decoding is governed by its weight distribution,
and can be well approximated by the union bound or variants
thereof~\cite{SS06}, especially at high SNRs. The ``truncated 
union bound'' is by far the simplest option, obtained by simply 
ignoring those terms~in the union bound for which the weight 
distribution is unknown. 
Consequently, it is neither 
an upper bound nor a lower bound.\linebreak
Nevertheless, we have found that in the high SNR regime,~it
pro\-vides a reasonable first-order approximation 
of performance under ML decoding for the codes at hand.
For example, \Fref{fig:union_boundPAC} shows the truncated union bound
for the two PAC~codes in Table\,\ref{table}
along with upper and lower bounds on their 
performance (under ML decoding)
obtained in simulations.

\hspace*{-0.72ex}Our results in this section also provide potential 
guidance~for the difficult problem of PAC code design.
Since both sequential decoding and list decoding achieve near-ML
performance, one important goal of rate-profiling 
should be to optimize the weight distribution at low weights.
The same criterion applies for the choice of the convolutional
precoder as well.
As~we~can\linebreak see from Table\,\ref{table}, 
the $(128,64)$ PAC
code with RM rate-profile\linebreak
succeeds at maintaining the minimum distance $d=16$ of the
self-dual Reed-Muller code, while ``shifting'' most of the codewords
of weight $16$ to higher weights. This 
establishes another reason for the remarkable performance of this code.

\vspace{2.70ex}
\section{Conclusions and Discussion}
\vspace{0.36ex}
\label{sec:last}

\noindent 
\looseness=-1
In this paper, we first 
observe that \Arikan's PAC~codes can~be 
regarded as polar codes with dynamically frozen bits and then, using 
this observation, propose an efficient list decoding algorithm 
for PAC codes. We show that replacing sequential~deco\-ding of PAC
codes by list decoding does not lead to degrada\-tion in performance,
providing the list size is sufficiently large.\linebreak
We then carry out
a qualitative complexity analysis of~the~two approaches, which
suggests that list decoding may be advanta\-geous in terms of 
worst-case complexity. We also study~the~per\-formance 
of PAC codes (and other codes) 
under ML~decod\-ing by estimating the first few terms in their
weight distribution. The results of this study provide constructive 
insights~into the remarkable performance of 
PAC codes at short blocklengths.~~

\looseness=-1
Based upon our results in this paper, we believe further~complexity
analysis of both sequential decoding and list decoding of PAC codes 
is warranted, including implementations in hardware.\ We hope our
work stimulates research in this direction.~~~

Finally, we would like to point out two important (and inter\-dependent)
but difficult questions regarding PAC codes that remain open.
What is the best choice of the rate profile? What is the best choice
of the convolutional precoder? We hope our results in \Sref{sec:analysis}
will contribute to further study of~these~pro\-blems. In turn,
effective resolution of these problems should make it possible 
to replicate the success of PAC codes at length $n = 128$ for
higher blocklengths.

\vspace{1.80ex}
\section*{Acknowledgement}
\noindent 
We are grateful to Erdal \Arikan\ for tremendous help with this work. 
Specifically, we are indebted to him for sharing the data and the source 
code of his sequential decoding program.
\vspace{0.81ex}


\end{document}